# SKYRMIONS IN THE SKYRME MODEL REVISITED

**NGUYEN DUY KHANH AND NGUYEN AI VIET**

Information Technology Institute, Vietnam National University, Hanoi, Vietnam

**ABSTRACT**

The possible boundary value problems of the skyrmions are reexamined in the original Skyrme model with two new viewpoints. The first one suggests to replace the infinity boundary point by the finite radius R, which is related to the density of the surrounding nuclear matter. The second viewpoint suggests to combine the global analysis with the numerical computation to give better results by avoiding the errors. It is interesting that the half integer topological skyrmions do exist in the model. Finally, the static properties of all possible skyrmions are calculated with the model parameters $F_\pi$ = 186 MeV, $m_\pi$ = 138 MeV and 1.57 < e < 4.17. The baryon masses are found to be too high compared to the experimental values, while the mass splitting between them is too small. On the other hand, the isoscalar radii and the axial coupling constant can be fitted with the experimental data with an error of 20% within the Balachandra's bound.

## I. INTRODUCTION

In the large $N_c$ limit, QCD can be regarded as an effective theory of mesons. Witten has argued that baryons might be modelled as the solitons without any further reference to the quark content in these theories [1]. However, it was Skyrme who has shown long ago for the first time that nucleons can be treated as solitons in the non-linear sigma model [2]. The interest toward the Skyrme model has revived after Adkins, Nappi and Witten [3,4] have successfully derived the static properties of nucleons from the model within 30% of the experimental values. Since only few first terms of the effective meson theory are known from the experimental data, one can modify the higher order terms in the original Skyrme model to improve the predictions in the modified Skyrme models [5,6] or in the extended Skyrme models [7,8].

In the Skyrme type models, the skyrmions are derived from the solutions of a singular non-linear differential equation, which has strange behaviors near the singular points in the numerical computations. It is difficult to judge whether these behaviors are the properties of the solutions or just computational errors. The purpose of this paper is two-fold. First, the global properties of the solutions will be studied systematically. By combining these properties with the numerical methods one can improve the computational results. Secondly, the new skyrmions, which are finite energy solutions, are found. In particular, the skyrmions which are restricted in a sphere with a finite radius R, will be examined in details.

The conservation of the topological charge of skyrmions leads to the boundary conditions of the differential equation of the model. Traditionally, skyrmions are treated as isolated solutions which extend their determination domains to infinity and vanish there. In a media, where the skyrmions must interact with each other or with the media, such a condition is not easy to fulfill, even not necessary to keep. Instead, one might consider the solutions, which are restricted in a sphere of a finite radius R. This radius in some sense can be related to the nuclear density of the skyrmion's surrounding media.

In a more careful analysis, we have found that the optimistic phenomenological results of the traditional skyrmions have some problems. For example, in the massless pion case, some integrals become infinite. As a consequence, the mass splitting between the nucleon and delta particles become zero. On the other hands, the magnetic isoscalar radius and the axial coupling constant are divergent. The finite results of [3] can be obtained if those integrals are artificially truncated. The introduction of the pion mass would have fixed this divergency problem, but the phenomenological predictions are not too optimistic. The classical masses are found too large and the mass splitting between the nucleon and delta particles is too small, while the predicted isoscalar radii and the axial coupling are reasonable.

Since the behaviors of the solutions in the Skyrme model are very complex, a more careful analysis is required. Our strategy is to carry out a lot of intensive numerical experiments to collect information about the behaviors of the solutions. Then with a global systematic analysis, we will try to explain the observed behaviors and improve the computation schema. In spite of the wide application field of the skyrmions ranging from nuclear and subnuclear physics to condensed matter physics, one knows relatively little about the global properties of the solutions in the Skyrme model. This paper classifies the skyrmions into three types. Beside the traditional skyrmions of Type I, the other two types are restricted in a finite sphere of radius R. The numerical computations based on a global analysis have shown that the new skyrmions do exist.

The paper is organized as follows: Sect.II. gives a brief overview about the Skyrme model and the possible boundary conditions for general solutions and skyrmions. The new skyrmions, which are restricted in a finite sphere of radius R, are also suggested. The values of R are shown to be related to the nuclear density. In Sect.III, the numerical results are presented for each type of skyrmion. The real skyrmions are limits between certain solution groups. By combining the global analysis with the numerical methods, one can improve the performance of the numerical computations and explain the behaviors of the solutions. Sect. IV presents the numerically computed static properties of various skyrmions. Sect.V summarizes the results and discusses the possible relevance of the new skyrmions in physics.

## II. THE MODEL

### II.1 Generalities

This paper investigates the skyrmions of the original Skyrme model with the massive pion [4], which is the simplest effective theory of the nonlinear pion field

$$U(\vec{x}, t) = \exp(i\pi_a(\vec{x}, t)\sigma^a), \tag{1}$$

where U is a SU(2) matrix function, $\pi_a(\vec{x}, t)$ and $\sigma^a$ are the triplets of pion fields and Pauli matrices.

The basic element to construct the model is $U^+ \partial_\mu U$, which used to be splitted into two parts as follows

$$U^+ \partial_\mu U = i\left(\mathcal{D}_\mu \pi^a + \Gamma^a{}_\mu\right)\sigma^a \tag{2}$$

where

$$\mathcal{D}_\mu \pi^a = \frac{\vec{\pi}\partial_\mu \vec{\pi}}{\pi^2}\pi^a + \frac{\sin(2\pi)}{2\pi}\left(\partial_\mu \pi^a - \frac{\vec{\pi}\partial_\mu \vec{\pi}}{\pi^2}\pi^a\right), \tag{3}$$

$$\Gamma^a{}_\mu = \frac{\sin^2(\pi)}{\pi^2}\epsilon^{abc}\partial_\mu \pi_b \pi_c \tag{4}$$

$$\pi = \pi(\vec{x}, t) = \sqrt[2]{\pi_a \pi^a}$$

The Lagrangian governing the model is as follows

$$\mathcal{L}_{SK} = \frac{F_\pi^2}{16}\text{Tr}\left(\partial_\mu U \partial^\mu U^+\right) + \frac{1}{32 e^2}\text{Tr}\left[\partial_\mu U U^+, \partial_\nu U U^+\right]^2$$

$$+ \frac{F_\pi^2 m_\pi^2}{8}\text{Tr}(U-1) = \frac{F_\pi^2}{8}\left(\mathcal{D}_\mu \vec{\pi} + \vec{\Gamma}_\mu\right)\cdot\left(\mathcal{D}^\mu \vec{\pi} + \vec{\Gamma}^\mu\right) + \frac{1}{16 e^2}\left[\left(\mathcal{D}_\mu \vec{\pi} + \vec{\Gamma}_\mu\right)\times\left(\left(\mathcal{D}_\nu \vec{\pi} + \vec{\Gamma}_\nu\right)\right]^2 + \tag{5}$$

$$\frac{F_\pi^2 m_\pi^2}{8}(\cos(\pi(\vec{x}, t)) - 1)$$

where the pion decay constant $F_\pi$ = 186 MeV and the pion mass $m_\pi$ = 138 MeV. Balachandra et al [9] have shown that the model is consistent with the exprerimental pion scattering data, only if the parameter e satisfies the bound condition

$$1.57 < e < 4.17 \tag{6}$$

The model (5) has the following conserved topological current

$$B^\mu = \frac{\epsilon^{\mu\nu\rho\tau}}{24\pi^2}\text{Tr}\left(U^+ \partial_\nu U \, U^+ \partial_\rho U \, U^+ \partial_\tau U\right) \tag{7}$$

The spherically symmetric skyrmions are constructed from the hedge hog ansatz as follows

$$\pi^a = F(r) x^a \tag{8}$$

where $\infty > r = F_\pi e \rho \geq 0$ is a dimensionless variable, which is used instead of the radial coordinate $\rho$ in the polar reference frame $(\rho, \phi, \theta)$. F(r) is a real value profile function satisfying an equation of motion, which minimizes the solition energy formula derived from the model (5).

The equation of motion of the model is a second order non-linear ordinary differential equation as follows:

$$\left(\frac{r^2}{4} + 2\sin^2(F(r))\right) F''(r) = \frac{m^2 r^2}{4} \sin(F(r)) - \frac{r F'(r)}{2} + \sin(2 F(r)) \left(\frac{1}{4} + \frac{\sin^2(F(r))}{r^2} - (F'(r))^2\right) \tag{9}$$

Instead of the pion mass, the following dimensionless parameter is used

$$m = \frac{m_\pi}{F_\pi e} \tag{10}$$

Since the values of $F_\pi$ and $m_\pi$ are known, the model has only the free parameter m to fit the theoretical predictions with the experimental values. The parameter m must satisfy the Balachandra's bound condition (6) as follows

$$0.18 < m < 0.48 \tag{11}$$

Eq.(9) can be solved with a boundary or initial condition. Such a condition is determined from the topological charge conservation. The conserved topological charge derived from Eq. (7) in the hedge hog ansatz is as follows

$$B^0 = -\frac{2}{\pi} \int_{F(0)}^{F(R)} \sin^2(F(r)) dF = \frac{F(0) - F(R)}{\pi} = N \tag{12}$$

The solutions with the topological charge $B^0 = N$ must satisfy the following condition

$$F(0) - F(R) = N \pi \tag{13}$$

Traditionally, $R = \infty$, the determination domain of the profile function F(r) expands from r =0 to r = $\infty$. In this paper, we also consider the possibilities of the solutions restricted in a sphere of finite radius R.

## II.2 Boundary conditions and solutions

Eq.(13) is still not a complete initial or boundary condition for Eq.(9). In our numerical experiements, we have observe that the trajectories of Eq.(9) are attracted to the points F(0) = k$\pi$/2 with a deviation less than 10%. The attempts to increase the precision in the neighbourhood of r = 0 often result in instabilities. Fortunately, we are able to prove the following proposition

**Proposition**

*The possible values F(0) of the profile function F(r) satisfying Eq.(9) must be n$\pi$/2 with the possible asymptotic formulae as follows*

$$F(r) = (4k+1)\frac{\pi}{2} + \beta r - \frac{3}{4} m^2 r^2 - \frac{\beta}{24} r^3 + m^2 \frac{r^4}{24} + \frac{\beta}{640} r^5 - m^2 \frac{r^6}{1080} \tag{14}$$

$$F(r) = (4k+3)\frac{\pi}{2} + \beta r + \frac{3}{4} m^2 r^2 - \frac{\beta}{24} r^3 - m^2 \frac{r^4}{24} + \frac{\beta}{640} r^5 + m^2 \frac{r^6}{1080} \tag{15}$$

$$F(r) = n\pi + \beta r + \frac{\beta m^2 r^3}{12 (4 \beta^2 + 1)} \tag{16}$$

**Proof:** Let us introduce the parameter a and the function Y(r) as follows

$$a = \sin(F(0)) \tag{17}$$

$$Y(r) = F(r) - F(0), \tag{18}$$

We have the limit

$$\lim_{r \to 0} Y(r) = 0 \tag{19}$$

In the case $a \neq 0$, Eq.(9) have the following asymptotic form in the neighbourhood of r = 0

$$Y''(r) = \frac{m^2 r^2}{8 a} - \frac{r}{4 a^2} Y'(r) + \frac{a \sqrt{1-a^2}}{r^2} \tag{20}$$

Eq.(20) has the following analytic solution

$$Y(r) = C[2] + \frac{1}{4} a m^2 r^2 + a \sqrt{2\pi} \, C[1] \, \text{Erf}\left(\frac{r}{2\sqrt{2}\,a}\right) + \frac{1}{2a}\left(\sqrt{1-a^2} - 2a^2 m^2\right) r^2 \, {}_2F_2\left(1, 1; \frac{3}{2}, 2; -\frac{r^2}{8a^2}\right) - a\sqrt{1-a^2} \, \ln(r) \quad (21)$$

where C[1] and C[2] are two arbitrary integration parameters.

The Gauss error function Erf (x) is defined [11] as follows

$$\text{Erf}(x) = \frac{2}{\sqrt{\pi}} \sum_{n=0}^{\infty} \frac{(-1)^n x^{2n+1}}{n!(2n+1)} = \frac{2}{\sqrt{\pi}}\left(x - \frac{x^3}{3} + \frac{x^5}{10} - \frac{x^7}{42} + \frac{x^9}{216} + \ldots\right) \quad (22)$$

$$\text{Erf}(0) = 0 \quad (23)$$

The generalized hypergeometric function ${}_2F_2(1, 1; b, 2; x)$ is defined [12] as follows

$$_2F_2(1, 1; b, 2; x) = \sum_{n=0}^{\infty} \frac{n!}{(b)_n (n+1)!} x^n, \quad (24)$$

$$_2F_2(1, 1; b, 2; 0) = 1 \quad (25)$$

where the Pochhammer symbol $(b)_n$ is defined as follows

$$(b)_0 = 1; \quad (b)_n = b(b+1)\ldots(b+n-1) \quad (26)$$

Since $\lim_{r\to 0}\ln(r) = -\infty$, the condition (19) requires C[2] = 0 and $a = \pm 1$, which means that $F(0) = (2k+1)\pi/2$, if $F(0) \neq n\pi$.

Eq.(21) leads to the following asymptotic formula of the profile function F(r)

$$F(r) = (2k+1)\frac{\pi}{2} + \frac{1}{4} \sin\left((2k+1)\frac{\pi}{2}\right) m^2 r^2 + \sqrt{2\pi}\,\beta\,\text{Erf}\left(\frac{r}{2\sqrt{2}}\right) - \sin\left((2k+1)\frac{\pi}{2}\right) m^2 r^2 \, {}_2F_2\left(1, 1; \frac{3}{2}, 2; -\frac{r^2}{8}\right) \quad (27)$$

In the infinitesimal neigbourhood of r = 0, the error and generalized hypergeometric functions can be approximated as follows

$$\text{Erf}\left(\frac{r}{2\sqrt{2}}\right) \approx \frac{r}{\sqrt{2\pi}}\left(1 - \frac{r^2}{24} + \frac{r^4}{640}\right) \quad (28)$$

$$_2F_2\left(1, 1; \frac{3}{2}, 2; -\frac{r^2}{8}\right) \approx 1 - \frac{r^2}{24} + \frac{r^4}{1080} \quad (29)$$

From Eqs (28)-(29), we obtain the asymptotic formulae (14) and (15).

In the case a = 0, the following approximations can be used when r → 0

$$\frac{\sin(F(r))}{r} \to F'(0) = \beta; \quad \frac{\sin(2F(r))}{r} \to 2\beta; \quad \frac{1}{4} + \frac{\sin^2(F(r))}{r^2} - (F'(r))^2 \to \frac{1}{4} \quad (30)$$

$$\frac{r^2}{4} + 2\sin^2(F(r)) = \frac{r^2}{4}\left(1 + 8\frac{\sin^2(F(r))}{r^2}\right) \to \frac{r^2}{4}\left(1 + 8\beta^2\right) \quad (31)$$

where $\beta$ is a finite value number. Eq.(9) is now reduced to the following asymptotic form

$$Y''(r) = \left(1 + 8\beta^2\right)^{-1}\left((m^2 + 2) r \beta - 2\frac{Y'(r)}{r}\right) \quad (32)$$

Eq.(32) has the following analytic solution

$$Y(r) = r\beta + \frac{m^2 r^3 \beta}{12(1 + 4\beta^2)} \quad (33)$$

satisfying Y(0) = 0 and Y'(0) = F'(0) = $\beta$. Eq.(33) implies the asymptotic formula (16).

In the light of the Proposition, the possible boundary condition for Eq.(9) must read

$$F(0) = N\pi + n\pi/2 \; ; \quad F(R) = n\pi/2 \tag{34}$$

Therefore, the solutions of Eq.(9) can be classified into the following types

i) The solutions of Type I satisfy the boundary condition

$$F(0) = N\pi + n\pi/2 \; ; \quad F(\infty) = n\pi/2 \tag{35}$$

ii) The solutions of Type II satisfy the boundary condition

$$F(0) = N\pi + n\pi/2 \; ; \quad F(R) = n\pi/2 \tag{36}$$

iii) The solutions of Type III satisfy the boundary conditions

$$F(0) = N\pi + n\pi/2 \; ; \quad F(R) = n\pi/2 \; ; \quad F'(R) = 0 \tag{37}$$

In the numerical computations, the non-skyrmion solutions are often mixed with the skyrmion ones, because they are denser. In many cases, they might cause strange behaviors. In order to understand these behaviors and to eliminate their influences in the final results, one must examine all the solutions and determine their possible relations to the skyrmions.

## II.3 Energy finiteness and skyrmions

Skyrmions are the solutions of Eq.(9), which have a finite energy. The energy finiteness will impose further constraints on Eqs. (35), (36) and (37). The classical energy derived from the Lagrangian (5) is as follows

$$M = 4\pi \left(\frac{F_\pi}{e}\right) M_0 = 4\pi \left(\frac{F_\pi}{e}\right) \int_0^R \left( \frac{\sin^2(F(r))}{4} + \frac{(r F'(r))^2}{8} + \frac{m^2 r^2}{4}(1 - \cos(F(r))) + \sin(F(r))^2 \left[ \frac{\sin^2(F(r))}{2 r^2} + F'[r]^2 \right] \right) dr \tag{38}$$

The finiteness of Eq.(38) has two implications. Firstly, the finiteness of the last term in the limit $r \to 0$ implies that $\sin(F(r)) \to 0$ or n is an even integer in Eqs. (35), (36) and (37). Hence, for the skyrmions we have the following boundary value

$$F(0) = N\pi \tag{39}$$

In other words, Eq.(39) selects out the skyrmions from the infinite energy solutions with the condition $F(0) = (N+1/2)\pi$.

Secondly, for the solution of Type I, in the limit $r \to \infty$, the first term in the integrand must tend to zero to keep the energy finite, which means that $F(\infty) = k\pi$ for the skyrmions of Type I. The skyrmions of Type II and Type III do not have such a restriction. So, skyrmions are classified accordingly into the following three types:

i) The skyrmions of Type I must satisfy the boundary condition

$$F(0) = N\pi \; ; \quad F(\infty) = 0, \pi \tag{40}$$

In principles, one can allow $F(\infty) = k\pi$. However, Eq.(9) has a periodicity symmetry meaning that that if F(r) is a solution, then $F(r) + 2k\pi$ is also a solution. So, one can consider Eq.(40) without loosing any generality. In Sect III, only the skyrmions of Type I with the condition $F(\infty) = 0$ are found.

ii) The skyrmions of Type II can have an integer topological charge if

$$F(0) = N\pi \; ; \quad F(R) = 0, \pi \tag{41}$$

and a half integer topological charge if

$$F(0) = N\pi \; ; \quad F(R) = \pi/2, 3\pi/2 \tag{42}$$

The periodicity symmetry implies that one can consider Eqs.(41) and (42) without loosing any generality. In Sect III, all the integer and half integer topological charged skyrmions of Type II are found.

iii) The skyrmions of Type III exist only if $m \ne 0$ satisfying the boundary conditions

$$F(0) = N\pi \; ; \quad F(R) = \frac{\pi}{2} \; ; \quad F'(R) = 0 \tag{43}$$

The periodicity symmetry implies that one can consider the values $F(R) = k\pi/2$, $k = 0,1,2,3$ without loosing any generality. However, if $F(R) = k\pi$, Eq.(9) and $F'(R) = 0$ imply $F''(R) = 0$. By differenciating both sides of Eq.(9), we can prove that all higher derivatives of F(r) at the boundary point r = R vanish. So, the solutions of Type III with $F(R) = k\pi$ will be a constant function. There is no skyrmion in these cases.

If $F(R) = \pi/2$, Eq.(9) implies

$$F''(R) = m^2 / (1 + 8/R^2) \ge 0 \tag{44}$$

If m = 0, F''(R) = 0 and all higher derivatives of the profile function vanish at r = R. In the case m ≠ 0, the function F(r) has a minimum at r = R. The numerical computations have shown that the solution trajectories starting from the point F(R) = $\pi/2$ will go up to reach the points F(0)=N$\pi$ for N≥ 1.

If F(R) =$3\pi/2$, Eqs.(9) implies

$$F''(R) = -m^2/\left(1 + 8/R^2\right) \leq 0 \tag{45}$$

In the case of m = 0, the solution is also a constant function. In the case m ≠ 0, the function F(r) has a maximum at r = R. Eq.(9) has a mirror symmetry, which means that if F(r) is a solution then $2\pi$ - F(r) is also a solution. So, the solutions starting from F(R) = $3\pi/2$ will be the mirror image of the ones starting from the F(R) = $\pi/2$. They will have a negative topological charge. So, without loosing any generality, one can consider only Eq.(43) for the half integer topological charged skyrmions of Type III. The numerical computations in Sect III have shown that such skyrmions exist.

## II.4 Nuclear density and new skyrmions

The radius R of the new skyrmions of Type II and Type III must have a physical meaning to have phenomenological implications. In fact, we are able to establish a relationship between the value of R and the nuclear matter density. We will follow the nuclear matter density patterns of P.Hansen et al. [10] and find the values of R corresponding to those patterns.

First, let us discuss about the units of the parameters to be used in this paper. The input parameters $F_\pi$ and $m_\pi$ of the Skyrme model are measured in MeV. On the other hand, the radii and distant in dense nuclear matter are measured in fm (fermi). The conversion between MeV and fm is always possible by using the conversion constant 197.33 MeV x fm = 1. In this paper, the dimensionless variable r = $F_\pi e \rho$ is used, where the radial coordinate $\rho$ is measured in fm. The nucleon baryon radius $\rho_{N0}$ = 0.72fm, where 1fm = $10^{-15}$cm = 1/197.33 MeV$^{-1}$. Hence the dimensionless radius of nucleon $r_{N0}$ can be expressed as follows

$$r_{N0} = F_\pi e \rho_{N0} = 0.68 e \tag{46}$$

Applying the Balachandra's bound (6), we obtain the following bound for the dimensionless radius of nucleon

$$1.07 < r_{N0} < 3.87 \tag{47}$$

From Eq.(45) we obtain the following ratio

$$\frac{e}{r_{N0}} = 1.47 \tag{48}$$

The volume of the nucleon as a sphere of radius $r_{N0}$ is as follows

$$V_{N0} = \frac{4}{3} \pi \rho_{N0}^3 = \frac{4}{3} \pi \left(\frac{r_{N0}}{F_\pi e}\right)^3 \tag{49}$$

The saturate baryon density of nuclear matter $\varrho_{N0}$ is where nucleons begin to touch. That will be different from the density in the volume $V_{N0}$ by a factor of $\frac{\pi}{3\sqrt{2}}$, because it is the density of the equal nucleon spheres closed-packed.

$$\varrho_{N0} = \frac{\pi}{3\sqrt{2}} \frac{1}{V_{N0}} = \frac{1}{4\sqrt{2}} \left(\frac{F_\pi e}{r_{N0}}\right)^3 = 0.148 \left(\frac{e}{r_{N0}}\right)^3 = 0.47 \left(\text{fm}^{-3}\right) \tag{50}$$

The baryon density of the skyrmions of Types II and III can be calculated as follows

$$\varrho_{NR} = \frac{1}{4\sqrt{2}} \left(\frac{F_\pi e}{R}\right)^3 = 0.148 \left(\frac{e}{R}\right)^3 \tag{51}$$

Comparing Eqs (50) and (51), we obtain the ratio of the baryon density of the given nuclear matter to the saturated nuclear density as follows

$$\frac{\varrho_{NR}}{\varrho_{N0}} = \left(\frac{r_{N0}}{R}\right)^3,$$

which can be related to the value of R as follows

$$R = r_{N0} \left(\frac{\varrho_{N0}}{\varrho_{NR}}\right)^{1/3} = \frac{F_\pi e}{197.33} 0.72 \left(\frac{\varrho_{N0}}{\varrho_{NR}}\right)^{1/3} = \frac{0.5045}{m} \left(\frac{\varrho_{N0}}{\varrho_{NR}}\right)^{1/3} \tag{52}$$

P.Haensel et al. [10] has classified the states of the nuclear matter which are equivalent to the ones in neutron stars. Using Eq.(52), the values of R can be classified according to the corresponding nuclear density as follows:

i) The nuclear matter state in the inner core of neutron stars: $0.94 < \varrho_{NR} < 7.052$ (fm$^{-3}$). The corresponding range of R at various values of m are listed in Table 1a.

| m | 0.48 | 0.273 | 0.18 |
|---|---|---|---|
| R | $0.4265 < R < 0.8342$ | $0.7493 < R < 1.4668$ | $1.2364 < R < 2.2256$ |

**Table 1a.** The ranges of R corresponding to the inner core of neutron stars

ii) The nuclear matter state of the outer core of neutron stars: $0.235 < \varrho_{NR} < 0.94$ (fm$^{-3}$). The corresponding range of R at various values of m are listed in Table 1b.

| m | 0.48 | 0.273 | 0.18 |
|---|---|---|---|
| R | $0.8342 < R < 1.3243$ | $1.4668 < R < 2.3284$ | $2.2246 < R < 3.5314$ |

**Table 1b.** The ranges of R corresponding to the outer core of neutron stars

iii) The nuclear matter state of the inner crust of neutron stars: $0.141 < \varrho_{NR} < 0.235$ (fm$^{-3}$). The corresponding range of R at various values of m are listed in Table 1c.

| m | 0.48 | 0.273 | 0.18 |
|---|---|---|---|
| R | $1.3343 < R < 1.5701$ | $2.3284 < R < 2.7606$ | $3.5314 < R < 4.186927$ |

**Table 1c.** The ranges of R corresponding to the inner crust of neutron stars

iv) The nuclear matter state of the outer crust of neutron stars $\varrho_{NR} < 0.141$. The corresponding range of R at various values of m are listed in Table 1d.

| m | 0.48 | 0.273 | 0.18 |
|---|---|---|---|
| R | $R > 1.5701$ | $R > 2.7606$ | $R > 4.186927$ |

**Table 1d.** The ranges of R corresponding to the outer crust of neutron stars

In numerical calculations of the solutions of Type II and Type III, we will vary the value of R in the above ranges to see in each state of the nuclear matter, what kind of skyrmion exists.

# III. NUMERICAL RESULTS

## III.1 Behavior of the skyrmions of Type I

First let us focus on the tradition choice of $F(\infty) = 0$ in Eq.(40). When $r \to \infty$, in this case, Eq.(9) reduces to the asymptotic form

$$F''(r) = m^2 \sin(F(r)) - \frac{2}{r} F'(r), \tag{53}$$

which has the following asymptotic solution at large r

$$F(r) = C \frac{e^{-mr}}{r} \tag{54}$$

where C is an integration constant.

The backward shooting method starts with the initial values of the profile function F(r) and its first derivative F'(r) at the point r = R are given by the asymptotic formula (54). For a good approximation, one can choose R=20. By varying the value of C, one can reach a solution within a certain chosen error as shown in Fig.1 for both cases m = 0 and m ≠ 0.

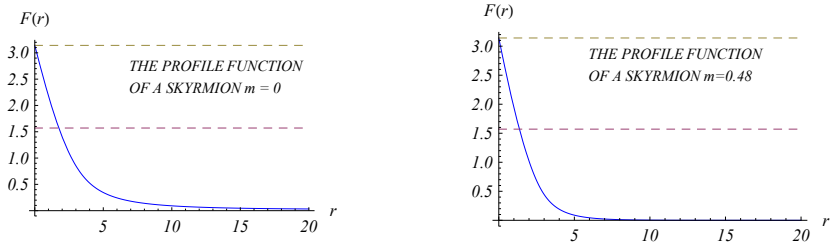

**Fig.1** The profile function of a skyrmion of Type I

It can be seen on Fig.1 that the skyrmion are shrunken under the attractive masss term.

It is worthy to note that if too high precision is required the solution becomes unstable and sometimes the shooting method cannot give a converged solution. Keeping in mind the results of the Proposition in Sect II., one can explain this observation as follows:

Let us draw the solution curve with m = 0 in Fig.1 with a magnification, in the neighbourhood r = 0 to the distant of $10^{-11}$ in Fig.2.

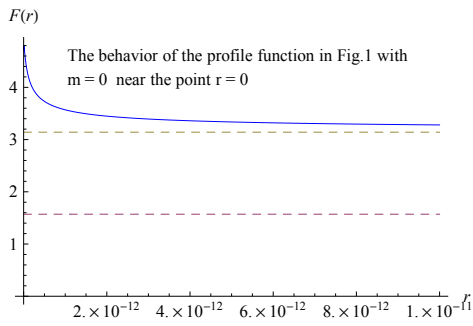

**Fig.2.** Magnification of the curve m = 0 in Fig.1 in the neighbourhood of r = 0.

The above curve goes very close to $F(0) = \pi$, but finally turns sharply to the points $F(0) = 3\pi/2$. So, the topological charge of the skyrmion derived from this curve is not 1. If a too high precision is required, it is very difficult to reach the point $F(0) = \pi$ by the backward shooting method. But such a solution still exists. We can demonstrate its existence by drawing all possible solution trajectories near the real solution as in Fig.3

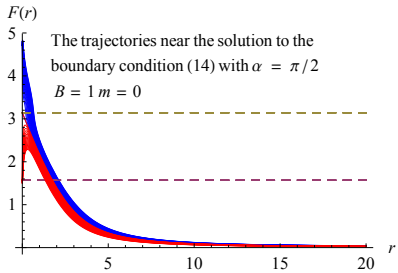

Since most solution trajectories of Eq.(9) swing between $F(0) = \pi/2$ or $F(0) = 3\pi/2$ just by a very small change in initial values of the parameter C. The trajectories can go as close to the point $F(0) = \pi$ as possible, but in the last step they turn to up or down direction sharply. However, there is a trajectory to seperate the upgoing and downgoing trajectories. That is the solution trajectory going to the point $F(0) = \pi$. As the up and down going trajectories are very dense near the point $F(0) = \pi$. In principles, one cannot reach this curve numerically with a very high precision. That is why one should use the asymptotic formula near the point $F(0) = \pi$, when accurate data is required.

In the case of $F(\infty) = \pi$ in Eq.(40), the solutions have complicated behaviors. As an example, the profile functions of a solution of Type I with the boundary condition $F(\infty) = \pi$, $F(0) = 2\pi$ and m = 48 are shown in Fig.6 with damping oscillating behavior.

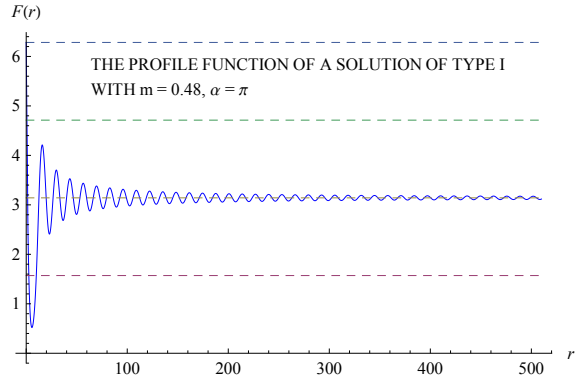

**Fig.6** The damping behavior of new skyrmions of Type I with a topological charge N=1

The results in Sect IV have shown that these solutions do not have a finite energy. So, the skyrmions of Type I exist only if $F(\infty) = 0$.

## III.2 Behavior of the skyrmions of Type II

Let us consider the choice of k=0 in Eq.(41) first. The backward shooting method is applied for the values of R corresponding to the different states in nucleon stars. So, the values of R = 1.2, 2.1, 2.7 and 3.5 are chosen. By varying the first derivative F'(R) at r = R, one obtains the solutions of B =1 at each value of R as in Fig.6.

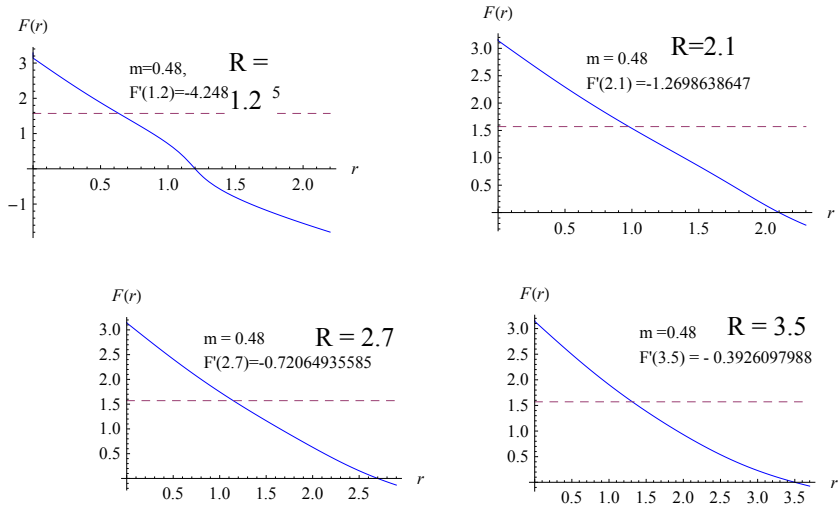

**Fig.6** The profile function of skyrmions of Type II with m = 0.48

So, it is possible to see that, the skyrmions of Type II with m = 0 and m ≠ 0 exist in the nuclear matter of all possible densities. The pion mass term has shrunk the size of the skyrmion of Type II with a small amount. We also observe the similar issue as in the case of the skyrmions of Type I. For example, the profile function with m = 0.18 looks like a solution of B = 1. In fact, it is a solution of B = 3/2 as shown in Fig.7

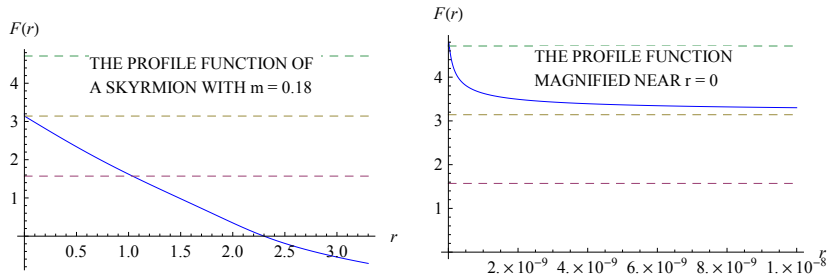

**Fig.7** The profile function of a skyrmion of Type II with m = 0.18

Similarly to the case of skyrmions of Type I, the close trajectories are drawn to see the existence of the "real" skyrmion of the baryon charge B = 1 in Fig.8.

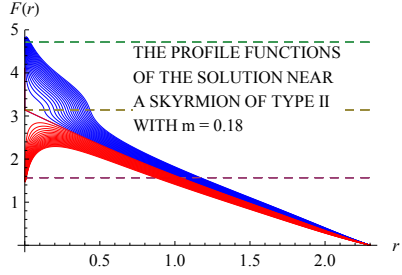

**Fig.8**. The profile functions of the trajectories of Eq.(9) near a skyrmion of Type II

The above numerical results explain why sometimes the shooting algorithm does not give convergent solutions, when a too high precision is required.

Since the numerically found solution trajectories of Type II starting from F(R) = 0 reach the values $\pi/2$, $\pi$, $3\pi/2$ before reaching the boundary points F(0) = N$\pi$, we can conclude that the skyrmions of Type II exist in all the choices of F(R)=0, $\pi/2$, $\pi$, $3\pi/2$ in Eqs.(41) and (42).

### III.3 Behavior of the skyrmions of Type III

In the case of the skyrmions of Type II, one can control F'(R) as the shooting angle. In the case the skyrmions of Type III, one can control only the parameters m and R in the shooting algorithm.

It is possible to observe that there are ranges of the parameter R and nuclear density where the solution of Type III of a certain integer topological charge. The transition points between the ranges of the integer topological charged solutions are the values of R where the half-integer skyrmions of Type III exist. We denote the values of R where the skyrmions of the topological (2k+1)/2 exist as $R_{(2k+1)/2}$. The corresponding nuclear densities are denoted as $\varrho^m_{NR(2k+1)/2}$ The values of $R_{(2k+1)/2}$ are found to be independent of the parameter m as listed in Table 2.

| Topological charge | m = 0.48 | m = 0.18 | m = 0 | Notation |
|---|---|---|---|---|
| N = 1/2 | 4.373523463472361 | 4.373523463472361 | 4.373523463472361 | $R_{1/2}$ |
| N = 3/2 | 5.427320890484484 | 5.427320890484484 | 5.427320890484484 | $R_{3/2}$ |
| N = 5/2 | 6.299784744755655 | 6.299784744755655 | 6.299784744755655 | $R_{5/2}$ |
| N = 7/2 | 7.054698399436239 | 7.054698399436239 | 7.054698399436239 | $R_{7/2}$ |
| N = 9/2 | 7.734095198154279 | 7.734095198154279 | 7.734095198154279 | $R_{9/2}$ |

**Table 2**. The values of the radial parameter R, where the skyrmions of Type 3 exist

By using Eq.(52), we can calculate the ranges of the nuclear density corresponding to the above values and ranges of R in Table 3.

| R | Density $\varrho_{NR}$ (fm$^{-3}$) for m = 0.48 | Density $\varrho_{NR}$ (fm$^{-3}$) for m = 0.18 | Topological charge | Skyrmion |
|---|---|---|---|---|
| < $R_{1/2}$ | > $\varrho^{0.48}_{NR\,1/2}$ | > $\varrho_{NR\,1/2}$ | 0 | No |
| $R_{1/2}$ | $\varrho^{0.48}_{NR\,1/2} = 0.006523859$ | $\varrho^{0.18}_{NR\,1/2} = 0.123711705$ | 1/2 | Yes |
| $R_{1/2} < R < R_{3/2}$ | $\varrho^{0.48}_{NR\,1/2} > \varrho_{NR} > \varrho^{0.48}_{NR\,3/2}$ | $\varrho^{0.18}_{NR\,1/2} > \varrho_{NR} > \varrho^{0.18}_{NR\,3/2}$ | 1 | No |
| $R_{3/2}$ | $\varrho^{0.48}_{NR\,3/2} = 0.003413835$ | $\varrho^{0.18}_{NR\,3/2} = 0.064736419$ | 3/2 | Yes |
| $R_{3/2} < R < R_{5/2}$ | $\varrho^{0.48}_{NR\,3/2} > \varrho_{NR} > \varrho^{0.48}_{NR\,5/2}$ | $\varrho^{0.18}_{NR\,3/2} > \varrho_{NR} > \varrho^{0.18}_{NR\,5/2}$ | 2 | No |
| $R_{5/2}$ | $\varrho^{0.48}_{NR\,5/2} = 0.00218284$ | $\varrho^{0.18}_{NR\,5/2} = 0.041393109$ | 5/2 | Yes |
| $R_{5/2} < R < R_{7/2}$ | $\varrho^{0.48}_{NR\,5/2} > \varrho_{NR} > \varrho^{0.48}_{NR\,7/2}$ | $\varrho^{0.18}_{NR\,5/2} > \varrho_{NR} > \varrho^{0.18}_{NR\,7/2}$ | 3 | No |
| $R_{7/2}$ | $\varrho^{0.48}_{NR\,7/2} = 0.001554403$ | $\varrho^{0.18}_{NR\,7/2} = 0.029476086$ | 7/2 | Yes |
| $R_{7/2} < R < R_{9/2}$ | $\varrho^{0.48}_{NR\,7/2} > \varrho_{NR} > \varrho^{0.48}_{NR\,9/2}$ | $\varrho^{0.18}_{NR\,7/2} > \varrho_{NR} > \varrho^{0.18}_{NR\,9/2}$ | 4 | No |
| $R_{9/2}$ | $\varrho^{0.48}_{NR\,9/2} = 0.001179697$ | $\varrho^{0.18}_{NR\,9/2} = 0.022370549$ | 9/2 | Yes |
| R > $R_{9/2}$ | < $\varrho^{0.48}_{NR\,9/2}$ | > $\varrho^{0.18}_{NR\,9/2}$ | > 5 | No |

**Table 3**. Values and ranges of the nuclear density corresponding to the solutions of Type III

In Sect II., we have seen that the integer topological charge solutions of Type III are not skyrmions. The transition points in the parameter R are where the skyrmions of half integer topological charge N+1/2 exist. The behavior of the profile function of Eq.(9) near the point F(0) =

$N\pi$ is similar to the cases of skyrmions of Type I and Type II discussed previously. The trajectories going to different boundary points at r = 0 are drawn in Fig.9

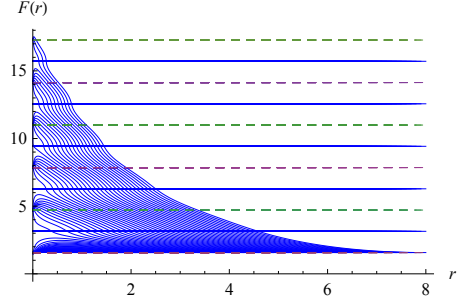

**Fig.9** The profile function trajectories of skyrmions of Type III with m = 0

The pion mass term does not change the transition points observed in the case m = 0, although it has shrunk the skyrmions by a very small amount. The existence of half integer topological charged skyrmions will be discussed in Sect. V.

# IV. STATIC PROPERTIES OF SKYRMIONS

In this section, the static properties of different skyrmions are computed from the numerical solutions of the profile function by using the input parameters of $F_\pi$ = 186 MeV and $m_\pi$ = 138 MeV. In fact, the traditional use of the nucleon and delta masses as inputs gives more impressive fits with experimental data. However, as shown below, since in the massless pion case the integrals become divergent, the good fits are obtained only by an artificial truncation. In the massive pion case, it is difficult to choose the masses as input parameters. The treatment of good fits will be discussed elsewhere using the modified models [16].

## IV.1 The formulae

Following the standard procedures of Refs. [2,3,4], one can use the following formulae to calculate the static properties of skyrmions. The topological charge now is identified as the baryon charge, when the skyrmions will be modeled as baryons.

### IV.1.1 Masses

For the massive pion m ≠ 0, the mass formulae for nucleon and delta can be expressed in terms of the model parameter m as follows

$$M_N = F_\pi/e\,(4\pi M_0) + 3\,e^3\,F_\pi/(4\pi\Lambda)*3/4 = 250.7\,m\,(4\pi M_0) + 0.006675/(4\pi\Lambda\,m^3); \quad (55)$$

$$M_\Delta = (4\pi F_\pi/e)\,M_0 + 3\,e^3\,F_\pi/(4\pi\Lambda)*15/4 = 250.7\,m\,(4\pi M_0) + 0.033375/(4\pi\Lambda\,m^3); \quad (56)$$

where the quantities $M_0$ and $\Lambda$ are given as the following integrals

$$M_0 = \int_0^R \left( \frac{\sin^2(F(r))}{4} + \frac{(r\,F'(r))^2}{8} + \frac{m^2\,r^2}{4}(1 - \cos(F(r))) + \sin^2(F(r))\left[\frac{\sin^2(F(r))}{2r^2} + (F'(r))^2\right] \right) dr \quad (57)$$

$$\Lambda = \int_0^R r^2 \sin^2(F(r))^2 \left( 1 + 4\left(\frac{\sin^2(F(r))}{r^2} + (F'(r))^2\right) \right) dr \quad (58)$$

For the massless pion m = 0, in the limit R → ∞, the asymptotic solution (54) give the integrand in Eq.(58) as follows

$$\lim_{r\to\infty} r^2 \sin^2(F(r))\left(1 + 4\left(\frac{\sin^2(F(r))}{r^2} + (F'(r))^2\right)\right) \to \lim_{r\to\infty}\left(r\sin\left(\frac{C}{r}\right)\right)^2 \to C^2 \quad (59)$$

which means that the integral $\Lambda$ is divergent when C ≠ 0 and the splitting between the nucleon and delta masses must be 0.

### IV.1.2 Isoscalar and magnetic isoscalar radius

The baryon charge density is derived from Eq.(7) as follows

$$\rho_B = -\frac{F'(r)\sin^2(F(r))}{r^2\,(2\pi^2)} \quad (60)$$

So, the isoscalar and magnetic isoscalar radii are defined as follows

$$<r_B^2> = \int_0^R r^2 \rho_B \, d^3x = -\frac{2}{\pi}(m/m_\pi)^2 \int_0^R r^2 \sin^2(F(r)) \, F'(r) \, dr \tag{61}$$

$$<r^2>_{m,I=0} = \int_0^R r^4 \rho_B \, d^3x \Big/ <r_B^2> = -\frac{\pi}{2}(m/m_\pi)^4 \int_0^R r^4 \sin^2(F(r)) \, F'(r) \, dr / <r_B^2> \tag{62}$$

The integral in Eq.(62) is divergent when m = 0 and R → ∞ for the same reason as in the case of Eq.(58).

### IV.1.3 The axial, pion-nucleon and pion-nucleon-delta coupling constants

The axial coupling constant has the following formula in the massive pion case

$$g_A = -\frac{D\pi}{3\,e^2} = -\frac{m^2}{3}\left(\frac{F_\pi}{m_\pi}\right)^2 D = -0.605545 \, m^2 \, D \tag{63}$$

where D is given by the following integral

$$D = \int_0^R \left(8 F'(r) \sin^2(F(r)) + 4 r F'(r)^2 \sin(2 F(r)) + \frac{4}{r}\sin(2 F(r))\sin^2(F(r)) + r^2 F'(r) + r \sin(2 F(r))\right) dr \tag{64}$$

The two last terms of the integral in Eq.(64) are divergent in the case m = 0 and R = ∞.

## IV.2 Static properties of the skyrmions of Type I

The m-dependency of the nucleon and delta masses is studied by computing all the masses with various values of the parameter m. In Table 4, the integrals Λ, $4\pi M_0$ and baryon masses are computed for the skyrmions of Type I with various values of the parameter m in the range 0.48 ≥ m ≥ 0.

| m | Λ | $4\pi M_0$ | C (R = 50) | $M_N$(MeV) | $M_\Delta$(MeV) |
|---|---|---|---|---|---|
| 0 | ∞ | 36.4613 | 0.259229455951825 | N/A | N/A |
| 0.001 | 58.53 | 36.4627 | 0.265735419232175 | 9085 | 45389 |
| 0.002 | 57.02 | 36.4627 | 0.272301601880600 | 1183 | 5840 |
| 0.0025 | 56.38 | 36.4628 | 0.275606715421500 | 626 | 3038 |
| 0.0027 | 56.14 | 36.4628 | 0.276932800793071 | 505 | 2428 |
| 0.003 | 55.8 | 36.4629 | 0.278926212835500 | 380 | 1790 |
| 0.005 | 53.97 | 36.4636 | 0.292343605705835 | 124 | 439 |
| 0.01 | 51.39 | 36.47 | 0.326786677596850 | 102 | 143 |
| 0.02 | 49.11 | 36.48 | 0.398829885382865 | 184 | 190 |
| 0.03 | 47.74 | 36.51 | 0.398829850000000 | 275 | 277 |
| 0.1 | 40.58 | 36.92 | 1.016853799491530 | 926 | 926 |
| 0.18 | 34.52 | 37.71 | 1.603985416872640 | 1702 | 1702 |
| 0.33 | 27.37 | 39.60 | 2.615546149747060 | 3276 | 3276 |
| 0.48 | 23.17 | 41.66 | 3.591290310000000 | 5012 | 5012 |

**Table 4.** The m parameter-dependency of the nucleon and delta masses as the skyrmions of Type I

C is the numerically found factor in the asymptotic formula (44) for R = 50, when the backward shooting method is used. The input parameters are the experimental values $F_\pi = 186$ MeV and $m_\pi$=138 MeV. The m-dependency of the nucleon and delta masses is interpolated from the above computed values and plotted in Fig.10

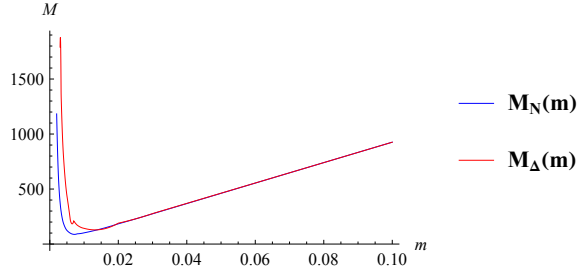

Fig 10. The m-dependency of masses of the skyrmions of Type I

From the value m = 0.035, the difference between the nucleon and delta masses becomes less than 1 MeV, while the masses grow linearly with m after reaching the experimental values of the nucleon mass at m = 0.1014 and the delta mass at m = 0.1321.

The nucleon mass reaches a minimum $M_N \sim 90$ MeV between 0.007< m < 0.008. The difference of the nucleon and delta masses reaches the experimental value $M_\Delta - M_N$ = 293 MeV at m = 0.005125. After the minimum, the masses grow again to reach infinity at m = 0. So, the delta mass reaches its experimental value again at m = 0.003419 and the nucleon mass does the same at m = 0.002167.

The static properties of baryons with the above critical values of the parameter m are computed and listed in Table 5.

| Static Properties | m = 0.48 | m = 0.18 | m = 0.1321 | m = 0.1014 | m = 0.005125 | m = 0.002167 | Experimental Values |
|---|---|---|---|---|---|---|---|
| $e$ | 1.55 | 4.12 | 5.616 | 7.317 | 145 | 342 | 1.57 < e < 4.17 |
| $M_N$ | 5012 MeV | 1702 MeV | **1232 MeV** | **939 MeV** | 120 MeV | **939 MeV** | 939 MeV |
| $M_\Delta - M_N$ | 0 | 0 | 0 | 0 | **293 MeV** | 3676 MeV | 293 MeV |
| $r_B^{I=0}$ | 1.08 fm | 0.495 fm | 0.376 fm | 0.295 fm | 0.0155 fm | 0.066 fm | 0.72 fm |
| $r_{B,m}^{I=0}$ | 1.46 fm | 0.71 fm | 0.549 fm | 0.436 fm | 0.0242 fm | 0.010 fm | 0.81 fm |
| $g_A$ | 2.65 | 0.456 | 0.255 | 0.154 | $7.76 \times 10^{-5}$ | < 0 | 1.23 |

**Table 5.** The static properties of baryons as the skyrmions of Type I

From the computed results, one can have some ideas about the actual predictive power of the Skyrme model. The isoscalar radii and the coupling constant $g_A$ can have good fits within the Balachandra's bound 0.18 < m < 0.48. For example, with m = 0.273, the values $r_B^{I=0}$ = 0.70fm, $r_{B,m}^{I=0}$ = 0.98fm and $g_A$ = 0.98 are obtained within 20% of the experimental values. However, the predicted masses are too large and the mass splitting is less than 0.4% MeV in the Balachandra's bound. The nucleon and delta masses approach their experimental values for the second time in the interval 0.002167 < m < 0.003419 with a reasonable mass difference, but the predicted values of the coupling constant $g_A$ and the isoscalar radii are too far from their experimental values. To improve the values of the masses, one must modify the model to have a smaller value of $M_0$.

## IV.3 Static properties of the skyrmions of Type II

The skyrmions of Type II are possible candidates of nucleons to be found in such a dense nuclear matter like the neutron stars. Let us choose the value m = 0.273, because the skyrmions of Type I have good fits with the experimental data of the isoscalar radii and the axial coupling constant at this value of m. We have calculated the static properties of the skyrmions of Type II with the topological charge N = 1 with at various values of R in Table 6.

| R | $M_N$ | $M_\Delta - M_N$ | $r_B^{I=0}$ | $r_{B,m}^{I=0}$ | $g_A$ | NOTES |
|---|---|---|---|---|---|---|
| N/A | 939 MeV | 293 MeV | 0.72 fm | 0.81 fm | 1.23 | Experimental Values |
| ∞ | 2648 MeV | 0.0035 MeV | 0.70 fm | 0.98 fm | 0.98 | Free skyrmions |
| 10 | 2661 MeV | 0.0036 MeV | 0.70 fm | 0.97 fm | 0.97 | Almost free skyrmions |
| 5 | 2726 MeV | 0.0043 MeV | 0.65 fm | 0.86 fm | 0.88 | Rare medium |
| 3.5 | 2875 MeV | 0.0051 MeV | 0.57 fm | 0.72 fm | 0.75 | Outer Crust of neutron stars |
| 2.7 | 3088 MeV | 0.0061 MeV | 0.49 fm | 0.61 fm | 0.64 | Inner Crust of neutron stars |
| 1.7 | 3800 MeV | 0.0084 MeV | 0.35 fm | 0.43 fm | 0.45 | Outer core of neutron stars |
| 1.2 | 4740 MeV | 0.011 MeV | 0.27 fm | 0.32 fm | 0.33 | Inner core of neutron stars |

**Table 6.** Static properties of the skyrmions of Type II with m = 0.273

The nucleon mass grow monotoneously with the increasing nuclear density. The isoscalar radii and the coupling constants decrease monotoneously with the increasing nuclear density.

## IV.3 Static properties of the skyrmions of Type III

The static properties of the skyrmions of Type III with a half integer topological charge are listed in Table 7 at various values of m

| Static Properties | $m = 0.48$ | $m = 0.273$ | $m = 0.18$ | $m = 0.1$ | $m = 0.05$ | $m = 0.01$ |
|---|---|---|---|---|---|---|
| $e$ | 1.55 | 2.72 | 4.12 | 7.42 | 14.84 | 74.19 |
| $M^{s=1/2}$ | 4963 MeV | 5556 MeV | 3504 MeV | 1898 MeV | 941 MeV | 199 MeV |
| $M^{s=3/2} - M^{s=1/2}$ | 0.0005 MeV | 0.002 MeV | 0.008 MeV | 0.047 MeV | 0.375 MeV | 47 MeV |
| $r_B^{I=0}$ | 1.162 fm | 0.382 fm | 0.166 fm | 0.051 fm | 0.013 fm | $0.51 \times 10^{-3}$ |
| $r_{B,m}^{I=0}$ | 2.004 fm | 1.139 fm | 0.751 fm | 0.417 fm | 0.208 fm | $41.2 \times 10^{-3}$ |
| $g_A$ | 2.881 | 0.343 | 0.065 | 0.006 | 0.0004 | $6.14 \times 10^{-7}$ |

**Table 7.** Static properties of the skyrmions of Type III with N=1/2

Let us note some qualitative properties the skyrmions of Type III with a topological charge N=1/2. These skyrmions exist at the nuclear density outside of the neutron stars. They are heavier than skyrmions of Type I. Their isoscalar radius and the axial coupling constant are smaller than the ones of skyrmions of Type I. However, their magnetic isoscalar radius is larger that the one of skyrmions of Type I.

## V. SUMMARY AND DISCUSSION

In this paper, we reexamine systematically the stable solutions with a conserved topological charge in the Skyrme model by a combined global and numerical method. The main results are summarized as follows:

1. In addition to the traditional skyrmions vanishing at infinity, the new skyrmions, which are restricted in a sphere of finite radius R, are also discussed and found.

    a. All the skyrmions of Type I have an integer topological charge and must satisfy F(0) = Nπ, F(R) = 0.

    b. The skyrmions of Type II can have an interger or a half integer topological charge.

    c. The skyrmions of Type III exist with a half integer topological charge only when pion mesons are massive. They exist with the condition F(0) = Nπ, F(R) = π/2.

2. The static properties of skyrmions show that in the Skyrme model the classical mass is too large leading to too small mass splitting between the nucleon and delta particle.

    a. Within the Balachandra's bound, the predicted baryon masses are too large compared to the experimental values. The mass splitting is almost zero within the bound.

    b. The predicted values of the isoscalar radii and the axial coupling constant can be fitted with 20% error with the experimental ones within the Balachandra's bound.

    c. Out of the Balachandra's bound, the masses and mass splitting can be fitted with the experimental values within an error of 50%. However, the predicted values of the isoscalar radii and the axial coupling contant will be unrealistically small.

3. The skyrmions of Type II with an integer topological charge can be used to model the nucleons in the nuclear matter

    a. The masses increase with the increasing nuclear density

    b. The isoscalar radii and the axial coupling constant decrease with the increasing nuclear density.

4. The existence of the half integer topological charged skyrmions needs further discussions for their physical relevance. The mass of the N=1/2 topological charged skyrmions is comparable and can be in some cases larger than the one of the N=1 topological charged skyrmions in the Skyrme model.

There are several physical implications of the above results. First, in order to use skyrmions as a model of baryons for a better fit with the experimental data, one must use a modified Skyrme model with a smaller classical mass. Instead of adding terms to the model, one must substract terms from the Lagrangian. In fact, as all the terms of the effective meson theories of QCD are not known, only the leading terms in these theories must be kept. The modified Skyrme models of [5,6] might be good candidates.

The idea of using the skyrmions restricted in a sphere of finite radius R is similar to the idea of using the Wigner-Seitz approximation for a continuous and homogeneous medium. However, phenomenological applications are possible only after the large mass issue will be solved.

# Acknowledgement

The work is supported in part by ITI-VNU and VIEGRID JSC. One of the authors (N.D.K.) is indebted to VIEGRID JSC and Madame Le Ngoc Hong for generous supports.

Nguyen Ai Viet is grateful to Professor I.Lovas for the discussions on the finite radius skyrmions.